\documentstyle[12pt,aaspp4]{article}
\received{~~} \accepted{~~}
\journalid{}{} \articleid{}{}

\begin{document}

\title{Testing the Blazar Paradigm: {\it ASCA} Observations of FSRQs with
Steep Soft X-ray Spectra}

\author{Rita M. Sambruna and Lester L. Chou} 
\affil{The Pennsylvania State University, Department of Astronomy and
Astrophysics, 525 Davey Lab, State College, PA 16802}

\author{C. Megan Urry} 
\affil{STScI, 3700 San Martin Dr., Baltimore, MD 21218}

\begin{abstract}

We present the first observations at medium-hard X-rays with {\it
ASCA} in 1998 August--November of four Flat Spectrum Radio Quasars
(FSRQs), characterized by unusually steep soft X-ray spectra (photon
index, $\Gamma_{0.2-2.4~keV} \sim 2-2.5$), as previously measured with
{\it ROSAT}. Such steep X-ray slopes are similar to those observed in
synchrotron-dominated BL Lacs and are unexpected in the context of the
recent blazar paradigm, where sources with strong emission lines (such
as FSRQs) are dominated in soft X-rays by a flat inverse Compton
tail. We find that the {\it ASCA} spectra of the four FSRQs are
consistent with a power law model with $\Gamma_{2-10~keV} \sim 1.8$,
flatter than their {\it ROSAT} spectra. This indicates the onset of an
inverse Compton component at energies $\gtrsim$ 2 keV, in agreement
with the blazar unification scheme.  However, these objects are still
anomalous within the blazar class for their steep soft X-ray continua
which, together with non-simultaneous data at longer wavelengths, hint
at the possibility that the synchrotron emission extends to soft
X-rays.  This would imply an anomalously high synchotron peak
frequency for a quasar with luminous broad lines, challenging current
blazar unification schemes. Alternatively, a plausible explanation for
the steep optical-to-soft X-ray continua of the four FSRQs is thermal
emission from the accretion disk, similar to the blazars 3C~273 and
3C~345. In the Appendix, we present fits to the SIS data in an effort
to contribute to the ongoing calibration of the the time-dependence of
the SIS response at low energies.

\end{abstract}

\section{Introduction}

Among radio-loud AGN, the subclass of blazars stands out for its large
luminosities and rapid variability timescales observed from radio to
$\gamma$-rays (Ulrich, Maraschi, \& Urry 1997; von Montigny et
al. 1995; Angel \& Stockman 1980). This can be readily explained as
non-thermal emission from a relativistic jet oriented at close angles
to the line of sight (Urry \& Padovani 1995; Blandford \& Rees
1978). While these continuum properties encompass the entire blazar
class, a further traditional subdivision is based on the strength of
the optical emission lines, with BL Lacertae objects (BL Lacs) having
weaker lines than Flat Spectrum Radio Quasars (FSRQs). However, since
the distribution of the line luminosity is continuous across the
blazar class (Scarpa \& Falomo 1997; Padovani 1992), the division
between BL Lacs and FSRQs is not bimodal but based on an arbitrary
equivalent width criterion.

More than twenty years after their discovery in the late 1970s, the
blazar ``zoo'' has been ordered according to the shape of the observed
spectral energy distributions (SEDs) from radio to $\gamma$-rays. The
SEDs appear to have a universal shape, with two distinct components
(Fossati et al. 1997; Sambruna, Maraschi, \& Urry 1996; Giommi,
Ansari, \& Micol 1995). The first component peaks anywhere from radio
to UV/X-rays and is interpreted as synchrotron emission from the jet
because it is highly variable, especially at higher energies, and
polarized (Ulrich et al. 1997). The second component extends to
$\gamma$-rays and its origin can be interpreted as inverse Compton
scattering of ambient photons off the jet electrons, either internal
(synchrotron-self Comtpon, SSC; Maraschi, Ghisellini, \& Celotti 1992)
or external (external Compton, EC; Ghisellini \& Madau 1996; Sikora,
Begelman, \& Rees 1994; Dermer, Schlickseier, \& Mastichiadis 1992),
although there are alternative scenarios, such as the hadronic models
(Protheroe \& Biermann 1997; Mannheim \& Biermann 1992). Depending on
the position of the synchrotron peak, blazars are classified as
High-energy peaked, or HBLs (peak in UV/X-rays) or Low-energy peaked,
or LBLs (peak in IR/optical).  An equivalent empirical division is
given by the ratio of the radio to X-ray fluxes, with $\alpha_{rx}
\lesssim 0.8$ in HBLs and $\alpha_{rx} \gtrsim 0.8$ in LBLs (Padovani
\& Giommi 1995). FSRQs have SEDs similar to LBLs and higher bolometric
luminosities (Padovani, Giommi, \& Fiore 1997; Sambruna 1997).

With the advent of deeper multicolor spectral surveys
(Laurent-Muehleisen et al. 1998; Perlman et al. 1998), it was quickly
realized that the distribution of synchrotron peaks is continuous and
not bimodal. FSRQs, LBLs, and HBLs can be arranged on a continuous
spectral sequence characterized by decreasing bolometric and emission
line luminosities, and increasing synchrotron frequency (Fossati et
al. 1998, 1997; Sambruna 1997; Sambruna et al. 1996). In addition, the
peak of the inverse Compton component appears to increase
proportionally to the synchrotron one along the same sequence, while
the ratio of the Compton to synchrotron luminosity decreases (Fossati
et al. 1998). On the basis of these trends a new scheme was proposed
where the transition from FSRQs to LBLs to HBLs is governed by a
change of a few physical parameters (Sambruna et al. 1996; Fossati et
al. 1998) and/or a change in the jet gaseous environs (Ghisellini et
al. 1998). However, selection biases especially at $\gamma$-rays need
to be addressed before reaching firm conclusions (Urry 1999).

According to this tentative new paradigm, blazars with strong emission
lines and/or thermal bumps (where the external radiation field is
strong) should be EC dominated and thus always have flat continua at
X-rays, where the inverse Compton tail dominates (Ghisellini et
al. 1998). At the other extreme are HBLs, where, as a consequence of
the high synchrotron peak frequencies, steep, convex (as a result of
the synchrotron losses), and highly variable X-ray spectra should be
observed. While recent observations with {\it ASCA} support this view
(Sambruna et al. 1999; Kubo et al. 1998), studies of larger samples of
FSRQs with {\it ROSAT} provided evidence for the existence of a
sub-class of ``X-ray-steep'' FSRQs (Padovani, Giommi, \& Fiore 1997;
Sambruna 1997), with soft X-ray photon indices $\Gamma_{0.2-2.4~keV}
\gtrsim 2$, similar to HBLs.  Moreover, recent multicolor blazar
surveys revealed that a sizable fraction (40\%) of the FSRQ population
exhibit $\alpha_{rx} \lesssim 0.8$ (Perlman et al. 1998), similar to
HBLs.

The existence of FSRQs with HBL-like spectra is unexpected in the
context of the blazar paradigm and requires an explanation. To gain
further insights we observed with {\it ASCA} four ``X-ray-steep''
FSRQs selected from our previous {\it ROSAT} study. The aim of our
{\it ASCA} observations is to measure the medium-hard X-ray continuum:
if synchrotron emission dominates the X-ray spectra of these systems,
like in HBLs, their {\it ASCA} spectra will be steep or even curved,
while if a Compton component is present (similar to most FSRQs) the
{\it ASCA} spectra will be flatter than at softer energies.  None of
the four targets is a known $\gamma$-ray emitter, contrary to most of
the FSRQs so far observed with {\it ASCA} (Kubo et al. 1998).

The structure of the paper is as follows. In \S~2 we discuss the
sample and the data, in \S~3 the results of spectral fits, and in \S~4
we briefly discuss the implications of our findings. Throughout the
paper, H$_0$=75 km s$^{-1}$ Mpc$^{-1}$ and $q_0=0.5$ are assumed.

\section{{\it ASCA} Observations} 

\subsection{The Sample} 

The targets of the present work are FSRQs with strong emission lines
selected from our previous {\it ROSAT} study (Sambruna 1997) according
to the following criteria: 1) they have steep X-ray spectra (photon
index, $\Gamma_{0.2-2.4~keV} \gtrsim 2$), as measured with the PSPC;
2) they are bright enough to obtain a high-quality {\it ASCA} spectrum
in a reasonable exposure time; and 3) they were never observed before
at medium-hard X-rays. These criteria gave a list of 7 FSRQs, four of
which were granted {\it ASCA} time during AO6.

The basic properties of the four FSRQs observed with {\it ASCA} are reported
in Table 1, where we list their names (column 1), redshifts (column
2), the observed radio flux densities at 5 GHz and the $V$ magnitudes
(column 3), and the Galactic column densities N$_H$ in the direction
to the sources (column 4). The latter were obtained from the 21 cm
maps of Elvis, Lockmann, \& Wilkes (1989) and Stocke et al. (1992),
and are accurate to $\pm 1 \times 10^{20}$ cm$^{-2}$ or better.  The
{\it ROSAT} slopes and 1 keV flux densities are summarized in columns 5 and
6, respectively, while columns 7--11 give the equivalent widths and
luminosities of the optical lines (in the sources' rest-frame), and
the composite spectral indices between radio (5 GHz) and optical (V
band), $\alpha_{ro}$, between the optical and X-rays (1 keV),
$\alpha_{ox}$, and between radio and X-rays, $\alpha_{rx}$. The
emission lines' parameters and the (non-simultaneous) multifrequency
fluxes were taken from the literature. None of the
sources in Table 1 was detected at $\gamma$-rays with EGRET (Hartman
et al. 1999) or other currently operating detectors. 

The composite spectral indices were evaluated using K-corrected flux
densities. For the K-correction, the fluxes were multiplied by
$(1+z)^{\alpha-1}$, where $\alpha$ is the energy index in the
appropriate band (F$_{\nu} \propto \nu^{-\alpha}$). In the radio,
individual slopes were used (K\"uhr et al. 1981 and Table 1).
In the optical, we used $\langle \alpha_{opt} \rangle =0.65$, the
average slope for a sample of radio-selected blazars (Falomo, Scarpa,
\& Bersanelli 1994). In the X-rays, the individual slopes from the
{\it ROSAT} PSPC observations were used (Table 1). 


\subsection{Data Reduction and Analysis}

We observed the four sources during {\it ASCA} AO6 in late 1998.  For a
description of the {\it ASCA} experiment see Tanaka, Inoue, \& Holt (1994).
In all cases, the Solid-State Imaging Spectrometers (SIS0 and SIS1)
operated in 1-CCD \verb+FAINT+ mode and the Gas Imaging Spectrometers
(GIS2 and GIS3) were used in Pulse-Height mode.  In order to apply
standard data analysis methods, the \verb+FAINT+ SIS mode was
converted into \verb+BRIGHT2+, applying the corrections for echo
effect and dark frame error (see ``The {\it ASCA} Data Reduction Guide'',
v.2, April 1997). Standard screening criteria were applied for the
data reduction, including rejection of the data taken during the
passage of the South Atlantic Anomaly and for geomagnetic cutoff
rigidity lower than 6 GeV/c.  We retained SIS data accumulated for
Bright Earth angles $>$ 20$^{\circ}$ and Elevation angles $>
10^{\circ}$, and GIS data accumulated for Elevation angles $>
5^{\circ}$. Only data corresponding to SIS grades 0, 2, 3, and 4 were
accepted.

The source spectra and light curves were extracted from circular
regions centered on the source position with radii of 4 arcmin for the
SIS and 6 arcmin for the GIS, which has a larger intrinsic point
spread function. The background was evaluated from blank-sky
observations in circular regions of the same size at the source's
position. We checked that larger extraction cells for the background
and/or different positions on the chip do not affect the final results
of our analysis. Details of the {\it ASCA} observations are reported
in Table 2, which lists: the date of the observation (column 2), the
net exposure after data screening (column 3), and the SIS0 and GIS2
net count rates in 0.6--10 keV and 0.7--10 keV, respectively.

Since no flux or spectral variability is apparent within the {\it
ASCA} exposures, the spectra were integrated over the entire duration
of the observations. The {\it ASCA} spectra were fitted using
\verb+XSPEC+ v.10. The SIS and GIS spectra were rebinned in order to
have a minimum of 20 counts in each spectral bin to validate the use
of the $\chi^2$ statistic. In order to increase the signal-to-noise
ratio we performed joint fits to the data from the four detectors,
leaving only the normalizations as independent parameters. The 1994
May response matrices were used for the GIS spectra, while for the SIS
data we used the matrices generated by the \verb+SISRMG+ program
(v.1.1, 1997 March).  The SIS and GIS data were fitted in the energy
ranges 0.6--10 keV and 0.7--10 keV, where the spectral responses are
best known.

\section{Spectral Fits and Results}

We first fitted the {\it ASCA} data using a single power law model,
modified at lower energies by the absorption column density N$_H$ (in
cm$^{-2}$). The Morrison \& McCammon (1983) cross-section for
photoelectric absorption was used, with the abundances of elements
heavier than hydrogen fixed at solar values. In the spectral fits
N$_H$ was left both free and fixed to the Galactic value (from Table
1). It must be borne in mind that systematic effects are present in
the two SIS detectors below 1--2 keV which lead to an overestimation
of the column density (Dotani et al. 1996) by at least $3 \times
10^{20}$ cm$^{-2}$, which has been increasing with the aging of the
mission (Yaqoob et al. 2000). Fits with free N$_H$ must thus be taken
with caution and are reported here for completeness only (see also
Appendix). No excess absorption was detected in the previous {\it
ROSAT} observations of the four blazars (Sambruna 1997).

The results from the fits to the {\it ASCA} data with a single power
law model are reported in Table 3. The power law photon index,
$\Gamma$, and the fitted N$_H$ are reported together with their 90\%
confidence errors ($\Delta\chi^2$=2.7 for one parameter of interest)
and the observed 2--10 keV flux, F$_{2-10~keV}$. Figure 1 shows the
residuals of the fit with a single power law plus fixed Galactic
absorption to the {\it ASCA} data of the four targets (solid symbols).
As apparent from Table 3, the fits with fixed N$_H$ are generally
acceptable, although the fits with free N$_H$ are improved, especially
for 0923+392. The 0.6--10 keV spectra of the four FSRQs are
characterized by relatively flat slopes, $\Gamma=1.7-1.8$, similar to
other FSRQs observed with {\it ASCA} (e.g., Kubo et al. 1998), and
flatter than their {\it ROSAT} PSPC slopes (Table 1). This strongly
suggests a different spectral component in the {\it ASCA} band,
although spectral variability between the epochs of the {\it ROSAT}
and {\it ASCA} observations can not be ruled out {\it a priori}.
However, we note that little or no spectral variability is observed at
both soft and hard X-rays for FSRQs repeatedly observed with {\it
ROSAT} and {\it ASCA} (Sambruna 1997; Kubo et al. 1998). Moreover, our
fits to the joint {\it ROSAT} and {\it ASCA} data (see below) are
entirely consistent with the two detectors sampling distinct regions
of the same continuum. We thus regard spectral variability as an
unlikely explanation of the different {\it ROSAT} and {\it ASCA}
slopes of our four FSRQs, and conclude that the broad-band X-ray
spectra of the sources is complex, and a soft excess is present below
$\sim$ 1 keV.


In the assumption that the steeper {\it ROSAT} slopes belong to a
different spectral component than {\it ASCA}, we attempted joint fits
to the {\it ASCA} and (non-simultaneous) {\it ROSAT} data in order to
characterize the shape of the soft excess.  For 0405--123, 0923+392,
and 1150+497, the {\it ROSAT} data lie above the extrapolation of the
{\it ASCA} power law at energies $\lesssim$ 1 keV (Figure 1, open
symbols).  We fitted the joint {\it ROSAT} and {\it ASCA} datasets of
these three sources adding to the hard X-ray power law one of the
following components to model the soft excess in the {\it ROSAT} band:
a thermal bremsstrahlung, a blackbody, and a second steeper power law
(double power law). During the fits, the column density was held fixed
to the Galactic value for all datasets, except for SIS0 and SIS1,
where it was left free to vary. The results of the fits with the three
models are reported in Table 4. It is apparent that all models provide
similar fits, and it is not possible to discriminate between a thermal
and non-thermal origin of the soft excess. There is, however, a slight
preference for the bremsstrahlung model for 0405--123
($\Delta\chi^2=5$ over the double power law model), and for a
non-thermal origin for 1150+497 (the double power law gives
$\Delta\chi^2=10$ over the thermal models).

The last column of Table 4 lists the observed 0.2--10 keV and 0.2--2.0
keV fluxes determined from the best-fit model.  The fraction of the
total (0.2--10 keV) flux in the soft component, i.e., the ratio of the
0.2--2 keV flux to the total 0.2--10 keV flux, is 12\% for 0405--123,
7\% for 0923+392, and 12\% for 1150+497.

In the case of 0736+017, the {\it ROSAT} plus {\it ASCA} data are well
fitted by a single power law, with a slope close to that derived from
the {\it ASCA} data only (Table 4). For this source, however, the {\it
ROSAT} data have large error bars, especially below 0.5 keV where the
absorption from the Galactic column density is large (Figure 1).  We
thus regard the evidence for a soft excess in this source as
tentative. 

In summary, we measured with {\it ASCA} the medium-hard X-ray continua
of four FSRQs characterized by steep 0.2--2.4 keV spectra and found
them consistent with single power laws with photon indices $\Gamma
\sim 1.8$. These are flatter than the {\it ROSAT} slopes and indicate
a soft excess in at least three cases (0405--123, 0923+392, and
1150+497). 

We are able to set interesting upper limits to the equivalent width
(EW) of an Fe K$\alpha$ emission line at 6.4 keV in the four FSRQs.
The Fe line is detected in other radio-loud AGN (Sambruna, Eracleous,
\& Mushotzky 1999) and in blazars like 3C~273 (Cappi et al. 1998;
Grandi et al. 1997). In our targets the Fe line would be redshifted in
the energy range 3--5 keV, close to the SIS sensitivity peak. We added
a narrow (width=0.05 keV, or FWHM $\sim$ 5600 km/s) Gaussian with
rest-frame energy 6.4 keV to the power law plus Galactic absorption
model and fitted the {\it ASCA} data. No fit improvement is obtained, with
the 90\% confidence upper limits on the Fe line EW of 47 eV for
0405--123 and 0923+392, 112 eV for 0736--017, and 81 eV for
1150+497. These are consistent with the upper limits found for
lobe-dominated radio-loud quasars (Sambruna et al. 1999) and
significantly lower than radio-quiet quasars of similar X-ray
luminosity (Nandra et al. 1997).

\section{Discussion and Conclusions}

We have observed with {\it ASCA} four FSRQs characterized by unusually steep
($\Gamma_{0.2-2.4~keV} \gtrsim 2$) X-ray spectra in the {\it ROSAT} band.
Our principal result is that all four targets have flat continua at
medium-hard X-rays described by photon indices $\Gamma_{2-10~keV} \sim
1.8$. Formally, the average slope for the four targets is $\langle
\Gamma_{2-10~keV} \rangle$ = 1.81 with 1$\sigma$ dispersion 0.05 (from
the fits with free N$_H$ in Table 3). This is consistent with the
average slope measured by {\it ASCA} for a group of 10 FSRQs,
$\Gamma_{FSRQs}=1.60$ with dispersion 0.14 (Sambruna et al. 1999). The
latter group include mostly sources with strong $\gamma$-ray
detections (Kubo et al. 1998), for which the flat X-ray continuum is
interpreted as the onset of the inverse Compton component peaking at
GeV energies, or high-$z$ sources where the flat Compton tail is
redshifted in the {\it ASCA} band. We thus conclude that our four
targets are also dominated by the onset of the inverse Compton
component at medium-hard X-rays.

The steep {\it ROSAT} spectra of the four targets, however, still
imply unusual properties. To clarify this point we assembled
(non-simultaneous) spectral energy distributions (SEDs) using our {\it
ASCA} and {\it ROSAT} data and longer wavelength fluxes from the
literature; the SEDs of the four sources are shown in Figure 2 (see
caption for references). The radio, optical, and X-ray data were
K-corrected following the procedure outlined in \S~2.1. For the IR
data, we used $\langle \alpha_{IR} \rangle$=1.27, from observations of
a sample of radio-selected blazars (Falomo et al. 1994). Although
poorly sampled in the intermediate wavebands, the SEDs in Figure 2
suggest large emission in the optical/IR bands\footnote{No correction
for the host galaxy emission was attempted; however, these sources are
all distant enough that the galaxy contribution in the optical band is
negligible.}  which connects smoothly to the {\it ROSAT} flux in most
cases. The SEDs have an upturn at medium-hard X-rays, where a
different component clearly begins.

Because of the limited {\it ROSAT} bandpass and the sparse sampling at
longer wavelengths in Figure 2, we can not entirely exclude a thermal
origin for the optical-to-soft X-ray flux. Indeed, in 0405--123 the
composite optical-to-UV spectrum shown by Corbin \& Boroson (1996)
appears steep with an upturn toward blue wavelengths, strongly
suggestive of a blue bump. Also evidence for a thermal bump in the SED
of 0923+392 is given by Elvis et al. (1994).  Viewed from this
perspective, the optical to soft X-ray emission of the four FSRQs of
the present study could be thermal emission from an accretion
disk. Indeed, thermal ``bumps'' at optical/UV have been observed for a
few blazars, including 3C~345 and 3C~273 (Smith et al. 1986; von
Montigny et al. 1997), and in 3C~279 during very low states (Pian et
al. 1999). In 3C~273, the high-energy tail of the thermal component is
detected with {\it ROSAT} and {\it SAX} (Leach, McHardy, \& Papadakis
1995; Grandi et al. 1997), with slopes $\Gamma_{soft} \sim 2.4-2.8$,
while a flat ($\Gamma_{hard} \sim 1.6$) continuum is measured at
energies $\gtrsim$ 2 keV (Grandi et al. 1997; Cappi et al. 1998;
Haardt et al. 1998).  The 0.2--2 keV component contributes $\sim$ 30\%
to the total 0.2--10 keV flux, larger than in our targets (\S~3). The
multiband continuum properties of 3C~273 are similar to our targets:
from its contemporaneous SEDs compiled in Von Montigny et al. (1997),
we derive K-corrected composite spectral indices $\alpha_{ro}=0.61$,
$\alpha_{ox}=1.3$, and $\alpha_{rx}=0.85$, well in the range of values
observed for our four FSRQs (Table 1). 


In the alternate hypothesis that the emission in the optical to soft
X-ray bands has a non-thermal origin, the most plausible candidate is
the high-energy tail of the synchrotron emission responsible for the
longer wavelengths. These objects would thus be similar to the
``intermediate'' LBLs 0235+164 and 0716+714, which have similarly
steep {\it ROSAT} spectra and flat X-ray continua at harder energies
(Urry et al. 1996; Madejski et al. 1996; Cappi et al. 1994; Kubo et
al. 1998) and synchrotron peaks at higher energies than in most
LBLs. The extension of the synchrotron component to soft X-rays in
0405--123, 0923+392, and 1150+497 (Figure 2) would indicate unusually
high electron energies and/or magnetic fields than in most FSRQs,
making these objects outliers of the general blazar trend outlined
above (\S~1).


The luminosities observed at the synchrotron peaks for our four FSRQs
are consistent with those measured for the two intermediate LBLs
mentioned above, 0235+164 and 0716+714 (Sambruna et
al. 1996). However, the quasars have more luminous MgII emission lines
by a factor $\gtrsim$ 10 than the two LBLs. From Table 1, L$_{MgII}
\sim (4-40) \times 10^{43}$ ergs s$^{-1}$ for the three quasars where
the MgII line was observed (0405--123, 0736+017, and 1150+497); for
the intermediate LBLs, L$_{MgII} \lesssim 4 \times 10^{42}$ ergs
s$^{-1}$ is measured for 0235+164 (Cohen et al. 1987), while 0716+714
exhibits a featureless optical spectrum (Stickel, Fried, \& K\"uhr
1993). This is in contrast with the unified paradigm of Ghisellini et
al. (1998), where the coexistence of sources with similar peak
frequencies and luminosities and largely ($\gtrsim$ 10) different
broad line luminosities is unexpected (e.g., their Figure 3).

In conclusion, our new {\it ASCA} observations of four FSRQs with
steep soft X-ray spectra are consistent with a flat continuum similar
to those of $\gamma$-ray-loud FSRQs, quite likely the onset of the
Compton component. The {\it ASCA} slopes are significantly flatter
than the {\it ROSAT} ones indicating a strong soft excess at low
energies in 3/4 sources. A possible origin for the latter is thermal
emission from an accretion disk (as in 3C~273 and 3C~445), and is
supported in at least two cases (0405--123 and 0923+392) by
independent longer-wavelength observations. Alternatively, the soft
X-rays are the high-energy tail of the synchrotron emission
responsible for the lower energies; in this case, their over-luminous
emission lines would present a challenge for current unification
schemes (Ghisellini et al. 1998). In either case, these sources
qualify as highly unusual in the blazar class; further broad-band
observations, especially in the critical optical-to-soft X-ray region,
are needed to determine the origin of the soft excess and the role of
these sources in the blazar class.

\acknowledgements

We acknowledge support from NASA contract NAS--38252 (RMS) and NASA
grant NAG5--2538 (CMU). An anonymous referee provided constructive
criticism which helped improving the paper.  This research made use of
the {\it {\it ASCA}} data archive at HEASARC, Goddard Space Flight Center,
and of the NASA/IPAC Extragalactic Database (NED) which is operated by
the Jet Propulsion Laboratory, California Institute of Technology,
under contract with the National Aeronautics and Space Administration.

\section{Appendix: Fits to the SIS data} 

It is well known that systematic effects are present in the two SIS
detectors below 1--2 keV which lead to an overestimation of the column
density (Dotani et al. 1996) by at least $3 \times 10^{20}$
cm$^{-2}$. The latter has been increasing with the aging of the
mission. A detailed study is under way (Yaqoob et al. 2000). In order
to help quantifying the magnitude of the effect, we performed spectral
fits to the SIS0 and SIS1 data of our sources, which were taken in
1998 August--November (Table 2).

The SIS0 and SIS1 spectra were fitted separately with an absorbed
power law, leaving the N$_H$ free to vary. The results are reported in
Table 5. Note that no excess absorption was detected with the {\it
ROSAT} PSPC in any of the sources (Sambruna 1997).  Comparing the
fitted N$_H$ from Table 5 to the value derived from the
(non-simultaneous) {\it ROSAT} observations, we find that the column
density measured with {\it ASCA} is always larger than the column
measured with the PSPC by $\Delta$N$_H \sim 6-12 \times 10^{20}$
cm$^{-2}$ for SIS0 and $\Delta$N$_H \sim 6-18 \times 10^{20}$
cm$^{-2}$ for SIS1.

\newpage 

\noindent{\bf Figure Captions}

\begin{itemize}

\item\noindent Figure 1: Residuals of the fits to the {\it ASCA} data
of the four FSRQs with a single power law plus Galactic
absorption. The fits were performed jointly to all the four {\it ASCA}
detectors, but only the SIS data are shown here for clarity (SIS0,
{\it asterisks}; SIS1, {\it filled triangles}). The best-fit model is
extrapolated into the {\it ROSAT} band and the non-simultaneous {\it
ROSAT} data are plotted for comparison ({\it open circles}). A strong
soft excess is apparent in 0405--123, 0923+392, and 1150+497, while
the {\it ROSAT} spectrum of 0736+017 is consistent with the {\it ASCA}
extrapolation.

\item\noindent Figure 2: Spectral energy distributions (SEDs) of the
four FSRQs using the {\it ASCA} data and non-simultaneous literature
fluxes at longer wavelengths. The SEDs are characterized by large
emission in the optical connecting smoothly to the soft X-rays.  Both
a thermal (from the accretion disk) and non-thermal (synchrotron
emission from the jet) origin plausibly account for the optical to
soft X-ray emission. Note the upturn at hard X-rays, marking the onset
of the Compton component. {\it References:} For the radio data:
Geldzahler \& Witzel (1981), K\"uhr et al. (1981), Shimmins \& Wall
(1973), Wright and Otrupcek (1990), Condon, Jauncey, \& Wright (1978),
Wall, Wright, \& Bolton (1976), Wills (1975), Owen, Spangler, \&
Cotton (1980), O'Dell et al. (1978), Owen et al. (1978), Owen et
al. (1980), Witzel et al. (1978), Owen, Porcas, \& Neff (1978), Genzel
et al. (1976), Patnaik et al. (1992).  For the IRAS data: Impey \&
Neugebauer (1988).  For the IR and optical data: Hyland \& Allen
(1982), Carballo et al. (1998), Wright, Mchardy, \& Abraham (1998).
The {\it ROSAT} data were taken from Sambruna (1997) and the {\it
ASCA} data are from this work.

\end{itemize}

\end{document}